%% file: main.tex
\documentclass[sigconf]{acmart}

\usepackage{xspace}
\usepackage{dsfont}
\usepackage[altpo, epsilon]{backnaur}
\usepackage{amsfonts,amsmath,amsthm}
\usepackage{listings}
\usepackage{tcolorbox}
\usepackage{mathpartir}
\usepackage{stmaryrd}
\usepackage{multicol}
\usepackage{multirow}
\usepackage[inline]{enumitem}
\usepackage{xspace}
\usepackage[dvipsnames]{xcolor}
\usepackage{float}
\usepackage{hyperref}
\usepackage{pifont}  
\usepackage{comment} 

\usepackage{cleveref} 
\usepackage{tikz} 

\renewcommand{\ttfamily}{\fontfamily{zi4}\selectfont\small}
\usepackage{etoolbox}
\AtBeginEnvironment{verbatim}{\small}

\usepackage{tabularray} 

\lstdefinelanguage{json}{
  morestring=[b]",
  morecomment=[l]{//},
  moredelim=[l][\color{black}]{:},
  moredelim=[s][\color{red}]{[}{]},
  stringstyle=\color{teal},
  keywordstyle=\color{black},
  commentstyle=\color{gray},
  showstringspaces=false,
}
\newcommand{\allnotes}[1]{}
\renewcommand{\allnotes}[1]{\textit{#1}}

\newcommand{\todo}[1]{\allnotes{{\bf\textcolor{red}{[#1]}}}}

\newcommand{\marg}[1]{\allnotes{\textcolor{magenta}{[MF: #1]}}}

\input{symb}

\begin{document}

\title{Hypergraph-Guided Regex Filter Synthesis for Event-Based Anomaly Detection}

\author{Margarida Ferreira}
\email{margarida@cmu.edu}
\orcid{0000-0002-1170-5124}
\affiliation{%
  \institution{Carnegie Mellon University}
  \country{USA}
}
\affiliation{%
  \institution{INESC-ID/IST}
  \country{Portugal}
}

\author{Victor Nicolet}
\email{victornl@amazon.com}
\orcid{0000-0002-3743-7498}
\affiliation{%
  \institution{Amazon}
  \country{USA}
}

\author{Luan Pham}
\email{luan.pham@rmit.edu.au}
\orcid{0000-0001-7243-3225}
\affiliation{
  \institution{RMIT University}
  \country{Australia} 
}

\author{Joey Dodds}
\email{jldodds@amazon.com}
\orcid{0009-0004-1534-6968}

\author{Daniel Kroening}
\email{dkr@amazon.com}
\orcid{0000-0002-6681-5283}

\affiliation{%
  \institution{Amazon}
  \country{USA}
}

\author{Ines Lynce}
\email{ines.lynce@tecnico.ulisboa.pt}
\orcid{0000-0003-4868-415X}
\affiliation{%
  \institution{INESC-ID/IST}
 \country{Portugal}
}

\author{Ruben Martins}
\email{ruben@cs.cmu.edu}
\orcid{0000-0003-1525-1382}
\affiliation{%
  \institution{Carnegie Mellon University}
  \country{USA}
}



\settopmatter{printacmref=false} 
\setcopyright{none}
\renewcommand\footnotetextcopyrightpermission[1]{}

\begin{abstract}
\input{abstract}
\end{abstract}

\maketitle

\input{sections/intro}
\input{sections/related_work}
\input{sections/motivation}

\input{sections/background}

\input{sections/algorithm}

\input{sections/similarity_score}
\input{sections/merge_regex}

\input{sections/merge-vertices}

\input{sections/evaluation}

\input{sections/conclusion}

\bibliographystyle{ACM-Reference-Format}
\bibliography{bibliography}

\newpage
\appendix
\input{appendix/extra}
\input{appendix/add-eval}
\end{document}

%% file: symb.tex
\definecolor{orange}{RGB}{245,100,0}
\definecolor{magenta}{RGB}{235,0,100}
\definecolor{blue}{RGB}{0,90,190}

\newcommand{\systemname}[0]{\textsc{HyGLAD}\xspace}
\newcommand{\zrhEvent}[0]{Out\xspace}
\newcommand{\numbaseline}[0]{7\xspace}

\newcommand{\refeq}[1]{Eq.~\ref{#1}}

\def\Hmat{\mathbf{H}}
\def\Verts{\mathbb{V}}
\def\Edges{\mathbb{E}}
\def\Graph{\mathbb{G}}

\newcommand{\regexL}[0]{\ensuremath{\text{UL}}\xspace}
\newcommand{\regexRC}[0]{\ensuremath{\text{RC}}\xspace}
\newcommand{\regexRCL}[0]{\ensuremath{\regexRC{}\text{L}}\xspace}
\newcommand{\regexRCLPlus}[0]{\ensuremath{\regexRCL^+}\xspace}

\def\throughput{{27~MB/s}\xspace}
\def\avgRecall{{{93\%}}\xspace}
\def\avgPrecision{{{81\%}}\xspace}

%% file: abstract.tex
\def\betterPrecision{{1.2}$\times$\xspace}
\def\betterRecall{{1.3}$\times$\xspace}


We propose \systemname, a novel algorithm that automatically builds a set of interpretable patterns that model event data. These patterns can then be used to detect event-based anomalies in a stationary system, where any deviation from past behavior may indicate malicious activity. The algorithm infers equivalence classes of entities with similar behavior observed from the events, and then builds regular expressions that capture the values of those entities. As opposed to deep-learning approaches, the regular expressions are directly interpretable, which also translates to interpretable anomalies. We evaluate \systemname against all \numbaseline unsupervised anomaly detection methods from DeepOD \cite{deepod2021} on five datasets from real-world systems.  The experimental results show that on average \systemname outperforms existing deep-learning methods while being an order of magnitude more efficient in training and inference (single CPU vs GPU). Precision is improved by \betterPrecision and recall by \betterRecall compared to the second-best baseline.


%% file: sections/intro.tex
\section{Introduction}

Event-based anomaly detection focuses on identifying irregularities in~\textbf{events}\footnote{https://opentelemetry.io/docs/specs/semconv/general/events/}, a type of machine-generated data that captures system activity in a predefined schema (e.g., time, actor, operation)~\cite{aws_cloudtrail_events, google_cloud_audit_logs,azure_activity_log}. These events are ubiquitous in cloud platforms and applications, where they serve as rich sources of data for understanding system behavior. Unlike unstructured logs~\cite{neurallog, du2017deeplog}, which contain free-text messages, events such as the one given in \Cref{fig:event-example} are suitable for systematic mining and modeling.

\begin{figure}[t]
\centering
\begin{lstlisting}[
language=json, 
basicstyle=\ttfamily\footnotesize, backgroundcolor=\color{gray!10},
frame=lrtb,
framerule=0pt,
]
{ "actor": {
    "id": "AttrService-InstanceRole-BTDN", 
    ...
  },
  "api": {
    "operation": "GetInstanceStatus",
    "request.data":
     {"instanceID": "i-12345", "asnDesc": "AMAZON-AES"}
  }
}
\end{lstlisting}
\vspace{-7pt}
\caption{Example of an event in OCSF schema~\cite{ocsf}.}
\label{fig:event-example}
\vspace{-1.4em}
\end{figure}

Recent studies \cite{he_survey_2021,he_empirical_2022,he_experience_2016,LANDAUER2023100470,le_log-based_2022,zhu_loghub_2023} report that despite interest in log- and event-based anomaly detection, several key challenges remain unresolved and block adoption. Because unsupervised approaches are less accurate than supervised approaches \cite{he_survey_2021, he_empirical_2022} and real-world labeled anomaly data is scarce, users must choose between generating labeled data for their unique system and relying on less powerful unsupervised techniques. In addition to accuracy, existing systems are not interpretable \cite{LANDAUER2023100470} or efficient \cite{he_survey_2021} enough for wide use. If an event is identified as an anomaly, it is crucial that this information is available to the user immediately, along with additional context to support further diagnosis and mitigation. Some systems generate billions of events per day, requiring fast and incremental detection techniques.

While deep learning approaches fit the requirements on accuracy~\cite{LANDAUER2023100470}, they are not interpretable and are computationally expensive. Deep learning approaches take an all-in view on anomaly detection where all of their processing is encoded in an uninterpretable statistical model. This gives them flexibility because they can be trained on almost any input, but the statistical models might be hiding many sub-problems in anomaly detection that can be tackled with efficient and interpretable algorithmic approaches. Our work identifies event categorization as a sub-problem that can be solved in an efficient and interpretable manner. In event categorization, the goal is to dynamically generate a set of general equivalence classes that covers the normal events in a system. 
Our approach is unconventional in that it solves the event categorization problem first and then uses the equivalence classes to detect anomalies: any event that does not fall into a known class is anomalous.
We show that the resulting anomaly detection approach is more accurate than existing deep-learning methods.

The key to building these equivalence classes is generalizing correctly over entities in the system. For example, imagine a system where three usernames appear in events, \texttt{System-Read-QCHXY}, \texttt{System-Read-RBLEA} and \texttt{System-Admin-DFGRD}. 
The two \texttt{System-Read} users can be observed to behave in the same way in event data.
If these two users fall into different equivalence classes, a third \texttt{System-Read} user that is created later will always appear as anomalous, even if it behaves the same as the other read accounts. The pattern we want is \texttt{System-Read-*}. On the other hand, if we generalize more, we run into another problem. The pattern \texttt{System-*-*} is too general, because it also captures our admin user, which can be observed behaving in ways that the read users don't.

We present \systemname, which synthesizes regular expression-based equivalence classes from a baseline of normal system behavior recorded in events. A key benefit of our regex-based equivalence model is that the model itself can be directly understood by humans. These benefits carry through to the anomaly detection method, which amounts to matching a few regexes per event (\throughput on a single CPU core), and where each anomaly is easily explained by comparing it to our known equivalence classes.

Our equivalence classes are a set of filters consisting of regular expressions, where each filter matches events. The regexes in the filters are a generalization of string values that appear in the events: they match at least those values and may additionally match other \emph{similar-looking} values representing entities that behave similarly. Generalizing solely based on the values themselves would introduce rules that accept potentially anomalous behavior.  The key insight of our approach is that events record the underlying \emph{relations} between entities in a system, and these relations can safely guide the generalization over the observed values. In other words, we combine and generalize entities that we observe acting in the same way with respect to other entities. We use this information to decide when to synthesize regexes that generalize over multiple~entities. 

In summary, we make the following contributions in this paper:
\begin{itemize}[nosep,leftmargin=*]
    \item We propose a new algorithm, \systemname, that builds interpretable regex-based equivalence classes for events using a graph-based similarity metric to group entities.
    \item We release three novel datasets with a total of 8.5 million events to benchmark event-based anomaly detection algorithms. Our datasets are collected from three large-scale cloud service systems containing thousands of resources and interacting actors, with labeled anomalous events.
    \item We extensively benchmark \systemname against seven state-of-the-art anomaly detection methods from DeepOD \cite{deepod2021} on five real-world datasets. Our results show \systemname is efficient and accurate: \systemname can scan \throughput of events on a single CPU core with \avgRecall recall and \avgPrecision precision.
\end{itemize}

%% file: sections/related_work.tex

\vspace{-10pt}
\section{Related Work}

Event-based anomaly detection is less explored than log-based detection. In this section, we review related studies and insights on their limitations that motivate our approach. 

\textit{Surveys.} Recent surveys \cite{he_survey_2021,he_empirical_2022,he_experience_2016,LANDAUER2023100470,le_log-based_2022,zhu_loghub_2023} highlight three key limitations of existing methods.  First, many high-performing techniques lack \emph{interpretability and explainability}, making it difficult for practitioners to trust or act on the outputs~\cite{he_survey_2021, he_empirical_2022, he_experience_2016, LANDAUER2023100470, zhu_loghub_2023}. Second, there is a \emph{heavy dependence on labeled training data}, despite anomalies being rare and difficult to define a priori in real-world systems~\cite{he_survey_2021, zhu_loghub_2023, LANDAUER2023100470}. Third, current approaches often fall short in \emph{efficiency and scalability}, particularly to the point of having the ability to identify anomalies in real time \cite{he_survey_2021,he_empirical_2022,he_experience_2016,le_log-based_2022,zhu_loghub_2023}.
 
\textit{DeepOD.} DeepOD \cite{deepod2021} is a comprehensive library of deep anomaly detection for tabular and time series data, focusing exclusively on deep learning-based methods. We compare our \systemname against all seven unsupervised anomaly detection methods implemented in DeepOD (Section~\ref{sec:baselines}), including DeepSVDD~\cite{deepsvdd}, DIF~\cite{dif}, RDP~\cite{rdp}, RCA~\cite{rca}, NeuTraL~\cite{neutral}, ICL~\cite{icl}, and SLAD~\cite{slad}. These methods were designed for general tabular anomaly detection, but we can apply them to event data by considering each row an event. DeepSVDD~\cite{deepsvdd} is a one-class classification method that learns to enclose normal data within a minimal-volume hypersphere in latent space. During inference, it computes the Euclidean distance from an event's embedding to the hypersphere center and flags high-distance samples as anomalies. DIF~\cite{dif} improves on an isolation forest by first projecting inputs through randomly initialized neural networks to capture complex feature interactions. At inference time, each event is passed through multiple random projections, and its anomaly score is the average depth across the resulting isolation. RDP~\cite{rdp} uses a Siamese network to align learned distances between samples with their distances in a random projection space, forcing the model to preserve global geometry. At test time, it detects anomalies that have a substantially larger distance than normal points on the learned representation. RCA~\cite{rca} trains an ensemble of autoencoders that selectively focus on reconstructing normal data patterns. During inference, the anomaly score is computed as the average reconstruction error across all autoencoders in the ensemble. NeuTraL~\cite{neutral} introduces a deterministic contrastive learning framework where a transformation network and an encoder are jointly trained to preserve identity under learned augmentations. 
ICL~\cite{icl} is a self-supervised method that uses feature masking and contrastive prediction to learn a one-class representation of normal instances. 
SLAD~\cite{slad} formulates anomaly detection as a proxy task called scale learning, where the model learns to distinguish subspace representations at different feature granularities. 
These methods share several limitations: (1) they require large volumes of training data, (2) require GPU for both training and inference, and (3) lack interpretability due to their black-box nature.


\textit{Log Parsing.} 
Log parsing techniques~\cite{he2017drain, huo_semparser_2023} convert logs into structured templates or equivalence classes (ECs), enabling anomaly detection by flagging log lines that deviate from known ECs. A widely used method, Drain \cite{he2017drain}, employs a fixed-depth parsing tree to efficiently extract log templates in an online setting. SemParser~\cite{huo_semparser_2023} advances this by incorporating semantic type inference for log parameters. However, a fundamental limitation shared by these approaches is that they fail to model the interactions between system entities (e.g., services, nodes, or users) that are implicitly encoded in the logs. As a result, they are insufficient for capturing higher-level behavioral patterns in complex systems. 

\begin{tcolorbox}[left=2pt,right=2pt,top=0pt,bottom=0pt,boxrule=0pt]
\textbf{Summary.}
Recent surveys highlight several key limitations in event-based anomaly detection: (1) lack of interpretability, (2) the need for unsupervised approaches, and (3) real-time detection. Meanwhile, log parsing methods such as Drain and SemParser offer unsupervised interpretable methods but fail to model interactions between system entities that are essential for understanding system behaviors. \textbf{We address these gaps by introducing \systemname, an unsupervised method that captures regular event patterns and entity interactions using historical event data. \systemname detects point-wise anomalies efficiently through interpretable patterns.}
\end{tcolorbox}

%% file: sections/motivation.tex
\section{Motivating Example}\label{sec:motivation}

We assume a system that generates events like the one shown in \Cref{fig:event-example}. Each event records an API call with information about the \texttt{"actor"} who executed it and the parameters (\texttt{"request.data"}). In the event in \Cref{fig:event-example}, a user with ID \texttt{AttrService-InstanceRole-BTDN} executes \texttt{"DescribeInstanceStatus"} on a computing instance with ID \texttt{i-12345} using the autonomous system (\texttt{ASN}) \texttt{"AMAZON-AES"}. Over time, the event history will show millions of events with different actors performing operations on varied~resources. 

\begin{figure}
\centering
\includegraphics[width=\linewidth]{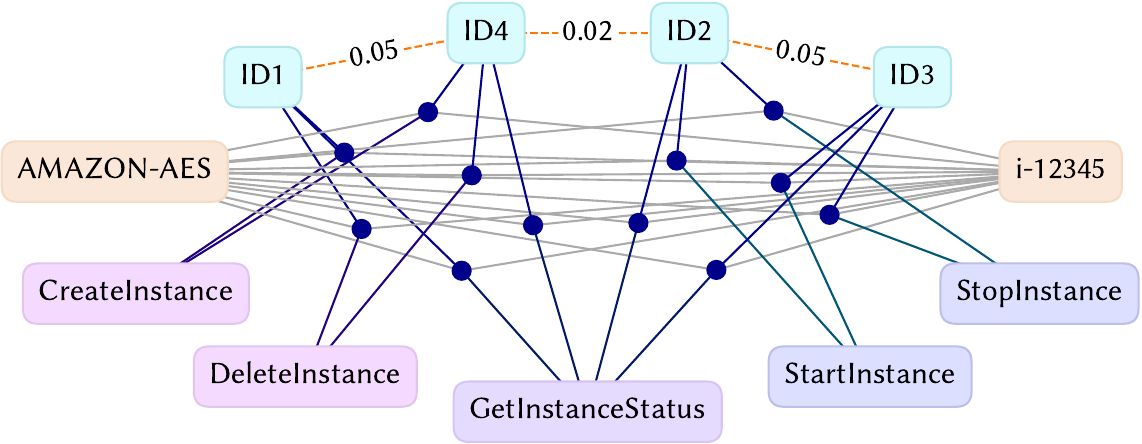}
\caption{A hypergraph showing 12 different events. Each event corresponds to a hyperedge represented as a dark-blue dot connecting all entities (represented as vertices) in the event.  The dark orange edges between ID vertices exemplify graph similarity between vertices. Pairs of vertices with more similar edge structures have a higher similarity value.}
\label{fig:all-events-graph}
\Description{A hypergraph diagram illustrating 12 different events, where each event is shown as a dark blue dot (a hyperedge) connecting several labeled nodes (vertices). The nodes include IDs (ID1 to ID4), operations (CreateInstance, DeleteInstance, GetInstanceStatus, StartInstance, StopInstance), and descriptors (AMAZON-AES, i-12345). A vertical red dashed line divides the IDs into two groups: the left group (ID1, ID4) is associated with creating and deleting instances, while the right group (ID2, ID3) is linked to starting and stopping instances. The hyperedges connect combinations of these fields to depict events.}
\vspace{-10pt}
\end{figure}

Our goal is to generate a \emph{model} consisting of general equivalence classes that cover the historical execution of the system. \systemname builds a hypergraph that represents the events, where each entity (e.g., user ID, computing instance) is a vertex, and each event is a hyperedge connecting two or more entities. The normal execution of our simplified example system includes 4 users operating on the same instance \texttt{i-12345} from the same ASN \texttt{AMAZON-AES}. \Cref{fig:all-events-graph} shows an example of a hypergraph built by \systemname for the system described above, using the following shorthand for user identifiers. 

{ \small
\begin{itemize}[leftmargin=0pt, label={},nosep]
    \item ID1  \(\gets\) \texttt{\textcolor{blue}{Attr}Service-\textcolor{magenta}{Instance}Role-\textcolor{orange}{BTDN}}
    \item ID2 \(\gets\) \texttt{\textcolor{blue}{Attr}Service-\textcolor{magenta}{Data}Role-\textcolor{orange}{QRIU}}
    \item ID3 \(\gets\) \texttt{\textcolor{blue}{Model}Service-\textcolor{magenta}{Data}Role-\textcolor{orange}{AUIB}}
    \item ID4 \(\gets\) \texttt{\textcolor{blue}{Model}Service-\textcolor{magenta}{Instance}Role-\textcolor{orange}{ZXWI}}
\end{itemize}
}
In this hypergraph, each hyperedge (i.e., event) connects to an identifier (light blue vertices on top), an operation name (purple vertices on the bottom), the ASN (\texttt{AMAZON-AES}), and instance \texttt{i-12345} (orange vertices on either side). Since all hyperedges connect to the same ASN and instance ID, the lines connecting all hyperedges to these are shown in a lighter gray color. The hypergraph edge structure highlights patterns in the observed events: IDs 1 and 4 connect to \texttt{CreateInstance} and \texttt{DeleteInstance}, but not \texttt{StartInstance} or \texttt{StopInstance}, whereas IDs 2 and 3 show the opposite pattern.

A naive way to model the behavior of this system is to keep a list of IDs that perform each of the observed operations. Then, for anomaly detection, accept an event as normal if that specific ID has executed that operation, otherwise flag it as anomalous. Although this solution correctly filters the observed events (identifying them as non-anomalous), it is too restrictive, which will result in false-positive anomaly reports. A good anomaly detector should generalize to also accept events \emph{like} those it has seen, not simply flag everything unobserved as~anomalous.

Information in the events is represented as strings so we use regular expression patterns of the observed strings to fix certain parts of entities and abstract away others. We can then efficiently match the regexes against values in new events for anomaly~detection.
The following regex \(r_0\) captures all the IDs in the events:
{ \small 
\begin{itemize}[leftmargin=0pt, label={}]
\item \(r_0\) = \texttt{\small{}\textcolor{blue}{[A-Za-z]\{8,11\}}Service-\textcolor{magenta}{[A-Za-z]\{4,8\}}Role-\textcolor{orange}{[A-Z]\{4\}}}.
\end{itemize}
}
\noindent
A more general anomaly detection filter would be to \emph{allow any ID matched by \(r_0\) to execute any of the five operations}. This would allow all the observed events but also others that are similar. However, this filter would be too permissive and match some events that should be considered anomalous. For example, ID2 has never deleted an instance, so it is possible that the user is not authorized to perform this operation; thus, an event that shows ID2 executing \texttt{DeleteInstance} should be flagged as anomalous.

Ultimately, we want a ruleset that maintains the \emph{relations between IDs and operations that we have observed}. The first step is to analyze the hypergraph and automatically extract the behavior patterns. \systemname extracts this information by measuring \emph{similarity} between vertices, taking into account the hypergraph structure of the events. The \emph{vertex similarity} between pairs of IDs (1,4), (4,2), and (2,3) is shown in the graph with dashed orange edges. (1,4) and (2,3) are more similar than (4,2), which matches the difference in behavior we noted previously. The algorithm progresses by merging vertices with similarity above a predefined threshold; in this case, it would merge the pairs (1,4) and (4,2). In doing so, an opportunity arises to generalize over their values and allow not only the same IDs but also similar values. 

To merge ID vertices (1,4), we need to build a regex that captures IDs like 1 and 4, but not like 2 and 3, so only 1 and 4 are allowed to execute \texttt{CreateInstance} and \texttt{DeleteInstance}. Given sets of positive and negative strings, \systemname's regex synthesis procedure outputs a regex that matches all the positive strings and none of the negative strings. Regex \(r_1\) below matches IDs 1 and 4 but not 2 and 3. On the other hand, \(r_2\) matches IDs 2 and 3 but not 1 and 4.
\begin{itemize}[leftmargin=0pt, label={}]
    \item \(r_1\) = \texttt{\small{}\textcolor{blue}{[A-Za-z]\{8,11\}}Service-\textcolor{magenta}{Data}Role-\textcolor{orange}{[A-Z]\{4\}}}
    \item \(r_2\) = \texttt{\small{}\textcolor{blue}{[A-Za-z]\{8,11\}}Service-\textcolor{magenta}{Instance}Role-\textcolor{orange}{[A-Z]\{4\}}}
\end{itemize}

\systemname's use of node similarity and regular expression synthesis allows it to generate patterns that keep the information relevant for characterizing behavior (the role of the identity in the naming scheme) and abstract over the rest (here, the generated suffix and the service name).
The ideal filter rules for this example~are:
\begin{itemize}[leftmargin=*,nosep]
    \item IDs that match \(r_1\) can execute \texttt{CreateInstance}, \texttt{DeleteInstance}, and \texttt{GetInstanceStatus}
    \item IDs that match \(r_2\) can execute \texttt{StartInstance}, \texttt{StopInstance}, and \texttt{GetInstanceStatus}
\end{itemize}

In this example, we showed the importance of considering the interaction between two types of entity, the operation name, and the user ID. These interactions are much more complex in real-world datasets that contain millions of events, each involving tens of entities. The hypergraph representation visible in \Cref{fig:all-events-graph} allows us to efficiently compress and manipulate large amounts of event data. \Cref{sec:definitions} formalizes the hypergraph model for events, and \Cref{sec:algorithm} presents \systemname's main algorithm.

%% file: sections/background.tex
\section{Events and Rules Representation}\label{sec:definitions}

Our model consists of a set of rules that match at least the events present in the training set. Rules and events are represented by similar objects: events are structures where attributes have values and some type, and rules are isomorphic structures where the values are regular expressions. A rule matches an event if they have the same attributes and, for a given attribute, the value in the event matches the regular expression in the rule. In this section, we formalize the concept of rules and events as one, keeping in mind that they only differ in the kind of values they contain.

Events contain entities with types and values:
\begin{definition}[Rule/Event]
A rule is a set of triples $(k,v,\tau)$ where $k$ is a key, $v$ is a value, and $\tau$ is the type of value $v$.
The keys in a rule must be unique.
\end{definition}

The signature of a rule or event is the set of its keys and types:

\begin{definition}[Rule Signature]
  The signature of a rule \\ $\{(k_i, v_i,\tau_i)\}_{1 \leq i \leq n}$, denoted by $\Sigma(e)$, is the set of its keys and types \\ $\{(k_i, \tau_i)\}_{1 \leq i \leq n}$.
\end{definition}

\begin{example}\label{ex:event-as-triple}
  \Cref{fig:event-example} shows an event in JSON format. The event is a rule that can also be represented as the following set of triples:
  
 \vspace{-1.6em}
 \begin{multline*}
 \small
      (\text{actor.id},
      \text{"AttrService-InstanceRole-BTDN"}, 
      \text{\textsc{Role}}),\\[-0.3em]
       \small
      (\text{api.request.data.instanceID},
      \text{"i-12345"},
      \text{\textsc{Instance}}),\\[-0.3em]
       \small
      (\text{api.operation},
      \text{"GetInstanceStatus"},
      \text{\textsc{EventName}})
  \end{multline*}
\end{example}

A rule $A$ matches another rule $B$ (usually, a concrete event) if they have the same signature and each of $B$'s values matches the pattern in rule $A$'s value for the same key.
Given a rule $e$ with signature $\{(k_i, \tau_i)\}_{1 \leq i \leq n}$, $e[(k_i,\tau_i)]$ denotes the value with key~$k_i$ and type~$\tau_i$. For example, for the rule in \Cref{ex:event-as-triple}, \(e[(\text{actor.id}, \text{\textsc{Role}})] = \text{"AttrService-InstanceRole-BTDN"}\).

\begin{example}
  The following is an example of a rule that accepts the event in \Cref{ex:event-as-triple}:

\vspace{-1.6em}
\begin{multline*}
\small
(\text{actor.id}, 
\texttt{"\footnotesize{}\textcolor{blue}{[A-Za-z]\{8,11\}}Service-\textcolor{magenta}{Instance}Role-\textcolor{orange}{[A-Z]\{4\}}"}, 
\textsc{Role})\\[-0.3em]
\small
(\text{api.request.data.instanceID},
\texttt{"i-12345"},
\textsc{Instance})\\[-0.3em]
\small 
(\text{api.operation}, 
\texttt{"GetInstanceStatus"}, 
\textsc{EventName})
\end{multline*}

\noindent
Its signature is the same as the signature of the event shown in \Cref{ex:event-as-triple}, and the regular expression matches the value in the event. The string values are otherwise equal.
\end{example}

To reason about the set of rules as a whole, we represent them in a hypergraph. Each vertex corresponds to an entity (a triple of attribute, value, and type) observed in the events. The hyperedges are the rules themselves, i.e., a set of entities.
We borrow the hypergraph formalism from \citet{dai_mathematical_2023}.

\begin{definition}[Rule Hypergraph]
A rule hypergraph $\Graph$ is defined by a set of vertices $\Verts = \{(k,v\tau)\}_{1 \leq i \leq n}$ and a set of hyperedges $\Edges$. The structure of the hypergraph can be represented as an incidence matrix $\Hmat \in \{0,1\}^{|\Verts| \times |\Edges|}$.
\end{definition}

%% file: sections/algorithm.tex
\section{Learning patterns from the hypergraph}\label{sec:algorithm}

\begin{figure*}
\centering
\vspace{-10pt}
\includegraphics[width=.9\linewidth]{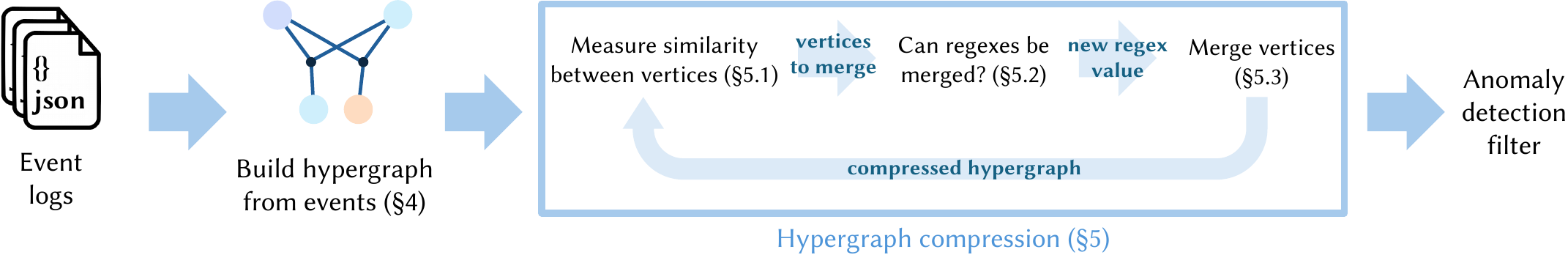}
\vspace{-.5em}
\caption{\systemname's design overview}
\label{fig:hyglad-design}
\vspace{-.5em}
\end{figure*}

In \Cref{sec:definitions}, we saw how \systemname builds the initial ruleset, which accepts exactly the events that have been observed, modeled as a hypergraph. In this section, we look into how the approach \emph{generalizes} this ruleset to accept similar events as well as those observed. \Cref{fig:hyglad-design} outlines \systemname's design. Once the hypergraph is built, we enter a loop where the algorithm merges similar vertices together until there are no possible merges left. In order to compute the similarity between vertices, we modify LSimRank~\cite{wu_lsimrank_2020} (see \Cref{sec:graph-sim}), which takes into account the graph structure in the similarity between two entities. To merge vertices with a similarity above a fixed threshold, we must synthesize a regular expression that captures both of their values (\Cref{sec:merge-regex}). The final step modifies the graph to replace the old vertices with the new, containing the new regex. The effect of this step on the graph can be computed in parallel with the previous; we explain in \Cref{sec:merge-vertex} how this can be used to extract negative examples for the regex~synthesis.

\begin{figure}
\centering
\includegraphics[width=\linewidth]{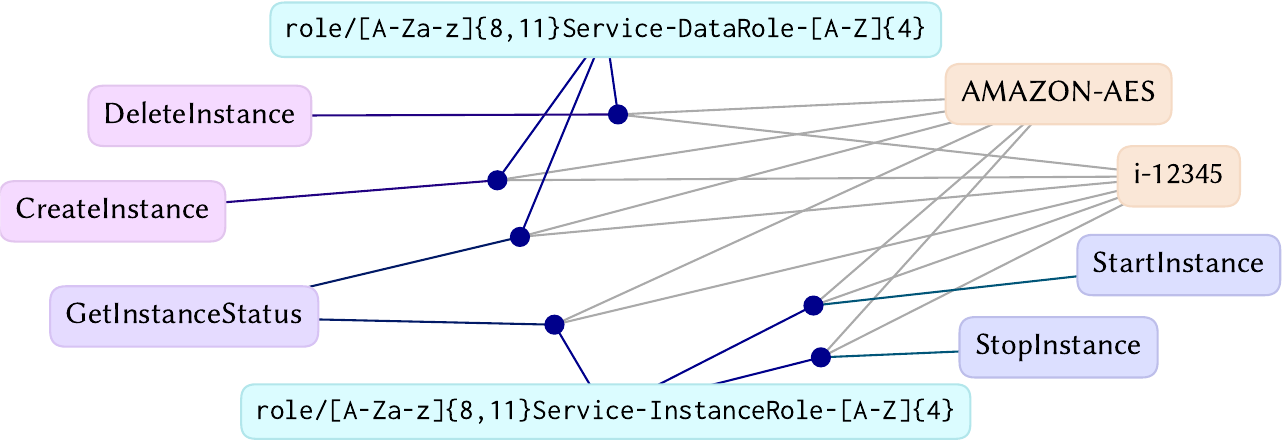}

\vspace{-.5em}
\caption{A hypergraph showing 6 rules.}
\label{fig:compressed-graph}
\Description{A hypergraph diagram representing 6 rules, where each rule is visualized as a dark blue dot (hyperedge) connecting a subset of entities (nodes). The nodes include operation types (CreateInstance, DeleteInstance, GetInstanceStatus, StartInstance, StopInstance), instance identifiers (i-12345), descriptor (AMAZON-AES), and two generalized ID patterns expressed as regular expressions:
1) [A-Za-z]{8,11}Service-DataRole-[A-Z]{4}
2) [A-Za-z]{8,11}Service-InstanceRole-[A-Z]{4}
The dark blue dots connect subsets of these entities, representing rules that match specific patterns of user actions.}
\vspace{-10pt}
\end{figure}

\begin{example}
Recall the example in \Cref{sec:motivation}: we aim to generalize the graph in \Cref{fig:all-events-graph} to allow IDs similar to 1 and 4 to execute CreateInstance and DeleteInstance, but not IDs similar to 2 or 3, as we have not observed them performing these operations. \systemname's method will infer the patterns shown before and merge IDs 1 and 4 into one single vertex, and IDs 2 and 3 into another, each represented by the respective regexes.
\Cref{fig:compressed-graph} shows the resulting graph. The ruleset in \Cref{fig:compressed-graph} is a generalization of the ruleset in \Cref{fig:all-events-graph}. All the events accepted by the ruleset in \Cref{fig:all-events-graph} are also accepted by the ruleset in \Cref{fig:all-events-graph}.

\end{example}

%% file: sections/similarity_score.tex
\subsection{Measuring Vertex Similarity}\label{sec:graph-sim}

To assess the similarity of two entities in the graph, we cannot simply examine the values alone; we must also use the relationships those entities have with others in the graph. SimRank~\cite{jeh_simrank_2002} uses the idea that two vertices are similar if they are connected to similar vertices. SemRank~\cite{milo_boosting_2019} and later LSimRank~\cite{wu_lsimrank_2020} modify the algorithm to consider the similarity between vertex labels in the calculation. In \systemname we modified LSimRank \cite{wu_lsimrank_2020} to work with hypergraphs and with the labels (event entities) needed.

\subsubsection*{LSimRank} 
LSimRank defines similarity between vertices in a graph as a matrix \(S\) defined as follows.
%
Let $\textit{In}(u)$ be the set of vertices connected by edges to $u$, 
$L(u,v)$ the label similarity between $u$ and $v$, 
and $c$ a \emph{decay factor} in $[0,1)$. 
Given an adjacency matrix $A$,
a label-similarity matrix $L$ with $L_{ij} = L(i, j)$, defined in \Cref{sec:label-sim}, and the normalization matrix $N$ such that
$N_{ij} = \begin{bmatrix}
  \frac{1}{|\textit{In}(i)| |\textit{In}(j)|} & \text{if} & i \neq j \\
  0 & \text{if} & i = j
\end{bmatrix}$, the similarity matrix $S$ is the limit of the following equation:

\vspace{-.9em}
\begin{equation}
  \label{eq:lsimrank-iter}
  S_0 = \mathds{1} \qquad
  S_{n+1} = \mathds{1} +  c \cdot L \circ N \circ A^\top S_n A
\end{equation}
where $\circ$ is the Hadamard product. Note that the sequence converges if $0 \leq c < 1$ \cite{wu_lsimrank_2020}.


\subsubsection*{Similarity score using LSimRank for hypergraphs} To adapt the similarity computation to hypergraphs, we apply a star expansion to obtain a graph; that is, we introduce a new vertex for each hyperedge and connect it to all the original nodes within that hyperedge. 
This approach retains the higher-order relations between entities through common events: entities should be similar if they appear in similar events. Since the vertices introduced by the star expansion do not have labels, we set the label similarity to be $L(u,v) = 0$ when $u \neq v$ and $u$ or $v$ is a hyperedge vertex, and $L(u,v) = 1$ when $u=v$.
The incidence matrix of the hypergraph $\Hmat$ is the adjacency matrix of the star expansion; thus, we replace $A$ for $\Hmat$ in \refeq{eq:lsimrank-iter}, and compute the similarity matrix~iteratively:
\[ S_0 = \mathds{1}_{|\Verts|+|\Edges|} \qquad 
S_{n+1} = \mathds{1}_{|\Verts|+|\Edges|} + c \cdot L \circ N \circ \Hmat^\top S_n \Hmat
\]

In practice, we set a limit $k$ on the number of iterations the algorithm goes through. The higher $k$, the larger the neighborhood of vertices that contribute to the similarity score. Because of the star expansion encoding, we must set $k > 2$ so that at least neighboring vertices are considered (as opposed to solely the hyper-edge vertex of the encoding). We experiment with different values of \(k\), and show the results in \Cref{sec:ablation}. 
Then, we define $\text{SimScore}(v_1, v_2, \Graph) = (S_k)_{v_1v_2}$.

\subsubsection*{Label Similarity in Typed Event Graphs}\label{sec:label-sim}

The similarity label~$L$ encodes purely local knowledge about the similarity between entities given by vertices in the graph, with no account for the graph structure.
Recall that each vertex label is a triple $(k, r, \tau)$ where $k$ is a key to a regex value $r$ of (semantic) type $\tau$ in some event.
Vertices that should not be merged because their values are too different or incompatible should score low in the similarity matrix. Thus, if vertices have different keys or types, we set the similarity to zero.


\paragraph{Similarity Between Regular Expressions} When the types and keys of two vertices match, the similarity of the label is computed from the values. In our setting, those values are regexes, which have a string representation. However, the distance between the string representations of the regexes is not a good measure of their similarity. The distance should take into account how close the sets of strings accepted by each regex are to each other, not the string representation of the regexes themselves. 
We adapt the Hausdorff set-distance metric \cite{hausdorff} to measure the distance between regexes: the distance between two regexes \( r_1 \) and \( r_2 \) is the Hausdorff distance between the sets of strings they match:
\begin{equation*}
d(r_1, r_2) = d_H(S_1, S_2) = \max \left( \sup_{s \in S_1}{\delta(s,S_2)}, \sup_{s' \in S_2}{\delta(s', S_1}) \right)
\end{equation*}
\noindent
where \( \delta(s, T) = \inf_{t \in T} d_s(s,t)\) computes the infimum of the string-distance between string \(s\) and the set of strings \( T\), given a string distance \(d_s\).
Since \(S_1\) and \( S_2 \) might be infinite, we compute the distance by sampling from the sets. 

%% file: sections/merge_regex.tex
\subsection{Regular Expression Synthesis}\label{sec:merge-regex}

To successfully merge two vertices, we need to compute a regex value that generalizes their values.
We implement a synthesis oracle \(\text{MergeRegex}\left(\{r_1, r_2\}, R^-\right)\) that returns a regex that matches all the words accepted by \(r_1\) and \(r_2\), and none of the words matched by the regexes in \(R^-\).
A trivial implementation of \(\text{MergeRegex}\) could simply consider the union \(r_1|r_2\), provided it does not intersect with any \(r^-\in~R^-\). The problem with this solution is that after successive executions \(r_1|r_2\) may grow arbitrarily large, and the intersection of two regexes may be exponential in their size \cite{succintregex}, making subsequent calls to \(\text{MergeRegex}\) and \systemname's inference of anomalies prohibitively~slow.

Instead of the union \(r_1|r_2\), \(\text{MergeRegex}\) outputs a generalization~\(r*\) such that all words matched by the union are also matched by \(r*\). At the same time, we ensure \(r*\) remains small by restricting our regexes to a specific subset of regular expressions sufficient to represent the strings encountered in events. 
The entities in events follow a precise structure. They correspond to IDs (user IDs, resource IDs), hashes, or date-time formats. These are strings that share constant prefixes, suffixes, or infixes interleaved with variable substrings and standard separators, like `-' or `/'. In the case of hashes, they typically consist of sequences of alphanumeric characters of a finite length. 

In our regular expression language, we consider two base types of regex units:
\begin{enumerate*}
    \item unions of string literals (\regexL{}s), which match one or more specific strings, and
    \item repeat character classes (\regexRC{}s), of the form \texttt{c\{m,M\}} with \texttt{m} \(\le\) \texttt{M} finite positive integers, which match any sequence of characters in character class \texttt{c} of length \texttt{m} to \texttt{M} (inclusive).
\end{enumerate*} 
We use the acronym \regexRCL to refer to a regex that is \emph{either} an \regexRC or an \regexL. Then, our language of regex filters consists of any concatenation of sequences of \regexRCL{}s, which we refer to as \regexRCLPlus. Since we also encounter strings with some portions omitted, the \regexRC{}s in \regexRCLPlus{}s can also be optional (regex operator \texttt{?}).

\begin{example}
    The regexes \texttt{i-} and \texttt{Owner|Maintainer|Evaluator}
    are \regexL{}s. The regex \texttt{[0-9]\{5,5\}} is a \regexRC{}. Thus, all regexes above are \regexRCL{}s, and \texttt{i-[0-9]\{5,5\}} is an \regexRCLPlus{}.
\end{example}

For any set of strings, it is always possible to efficiently compute \regexRCLPlus{}s that match all strings. Furthermore, for any two \regexRCLPlus{}s, it is efficient to compute a third \regexRCLPlus{} that includes their union. Finally, it is efficient to compute the intersection between two \regexRCLPlus{}s.

\begin{example}
    Consider the strings ``i-12345'', ``i-12739'', and ``i-12119''. The \regexRCLPlus{}s \texttt{i-(?:12345|12739|12119)}, \texttt{i-12(?:345|739|119)}, \\
    \texttt{i-12[0-9]\{3\}}, and \texttt{i-[0-9]\{5\}} are all valid \regexRCLPlus{}s that match all three strings.
\end{example}


The implementation of \(\text{MergeRegex}(\{r_1, r_2\}, R^-)\) starts by enumerating candidates \(r^+\) that are in our sublanguage of regexes and contain \(r_1|r_2\). From these candidates, we remove those that intersect any \(r^-\in~R^-\). These operations can be computed efficiently for the small regexes we include in our language.

Finally, to pick among valid candidates \(r^+\), we rely on a heuristic measure of \emph{cost} of an \regexRCLPlus to sort possible candidates. We consider two values for the cost function:
\begin{enumerate*}
    \item number of nodes in the regex (number of tokens and operations). Smaller regexes are desirable because they are more efficient to match, more readable, and more general than larger regexes. 
    \item number of words accepted by the regex. The generalization of regular expressions results in the filter accepting more strings. This generalization should come at a cost, since we are choosing to consider normal values that were not previously observed.
\end{enumerate*}
These two components of cost help the algorithm balance between saving observed strings as literals in a \regexL, making the filter more specific to the values it has observed, and generalizing them as a \regexRC, making the regexes more efficient and more general.
Computing this cost for arbitrary \regexRCLPlus{}s is very efficient, since the number of nodes and the number of accepted words are both finite and can be computed in linear time in the number of nodes of the regex.

%% file: sections/merge-vertices.tex
\subsection{Merging Vertices}\label{sec:merge-vertex}

When merging two vertices, we take into account the similarity between the hyperedges that are connected to those two vertices. The modification of the graph is straightforward, and we refer the reader to \Cref{sec:extra} for a formal description. A key insight is that during the merge operation, we try to maintain an invariant over the whole graph, and this is what allows us to extract negative examples for the regular expression synthesis problem.

\paragraph{Invariant: Rule Uniqueness}
To ensure the set of rules remains tractable, the algorithm maintains the invariant that the edges in the graph are unique: there are no two edges with the same signature (same keys and types) and intersecting values (e.g., when values are regexes).
This can be expressed in terms of how many edges (i.e., rules) match a given event:

\begin{proposition}[Edge Uniqueness and Event Matching]
  The edges in $\Edges$ are unique iff for any given event $\eta = \{(k_i, r_i, \tau_i)\}_{1 \leq i \leq n}$, there is at most one edge $e \in \Edges$ such that 
  $e$ and $\eta$ have the same signature and their values match.
\end{proposition}

When attempting to merge vertices $v_1$ and $v_2$, the set of negative examples $R^-$ is calculated by taking all the edges that are not in the immediate neighborhood of $v_1$ and $v_2$ but have the same signature, and extract the values corresponding to the same key and type in those edges that have the same values for keys that are not~$k_1$.


To fully describe the negative example extraction, we first define an inter-neighbor relation for a hypergraph.

\begin{definition}[Inter-neighbor Relation]
  The inter-neighbor relation $N \subset \Verts \times \Edges$ on a hypergraph $\Graph = (\Verts, \Edges)$ with incidence matrix $\Hmat$ is defined as
   \( N = \{(v,e) \mid \Hmat(v,e) = 1, v \in \Verts, e \in \Edges \} \).
\end{definition}

\begin{definition}[Hyperedge Neighbor Set]
  The hyperedge inter-neighbor set of vertex $v \in \Verts$ is defined as  \( N_e(v) = \{e \mid vNe, v \in \Verts, e \in \Edges \} \).
\end{definition}

 Note that by construction, it should be true that $r_1 \notin R^-$ and $r_2 \notin R^-$.
 Formally, let
 \begin{multline*}
   S :=  \{ e \mid e \in \Edges
   \;\wedge\; \exists e' \in N_e(v_1) \cup N_e(v_2) \cdot \Sigma(e') = \Sigma(e) \;\wedge \\
   \forall (k, \tau) \in \Sigma(e) \text{ s.t. } k \neq k_1, \tau \neq \tau_1 \cdot e[(k, \tau)] = e'[(k,\tau)] \}
 \end{multline*}
 be the set of edges that are not in the neighborhood of $v_1$ and $v_2$ but have the same signature as some edge in the neighborhood and differ from that edge only by the value at $k_1, \tau_1$.
 Then the set of negative examples is:
 \[
 R^- := \bigcup_{e \in S} \{  e[(k_1, \tau_1)] \}
 \]


%% file: sections/evaluation.tex
\section{Experiments}\label{sec:evaluation}

\subsection{Experimental Settings} \label{sec:experiments}

\subsubsection{Datasets} \label{sec:dataset} 
There are no open-source event datasets specifically designed for point-wise anomaly detection. We address this by releasing three novel datasets collected from real-world cloud systems: \textbf{Falcon}, \textbf{Flask}, and \textbf{Live}. Falcon is a cloud-based web application platform consisting of 363 unique actors and 138{,}292 resources. Flask is a microservice-based music catalog and retrieval system comprising 531 actors and 143{,}353 resources. Live is a real-time cricket scoring platform, serving 2{,}507 actors and managing 1{,}404 resources. Each system generates a large volume of events across multiple layers (e.g., API calls, configurations, and resource updates). We collect normal events over periods ranging from 7 days to 1 month for training, and 10 days of events including anomalies for~testing.

We further evaluate \systemname on \textbf{\zrhEvent{}}, collected during an incident from a global cloud provider. Customers were unable to log in to the Console in Region A for two hours. Around 33.1\% requests failed due to 4xx/5xx errors. The incident involved multiple teams and impacted several services. The event data includes more than twenty thousand unique actors and two hundred resources.

Additionally, we evaluate \systemname on the \textbf{BETH} dataset~\cite{highnam2021bethdata}, originally proposed for cybersecurity benchmarking. In BETH, anomalous events are considered `out-of-distribution', as they are generated from a data distribution not seen during training. Therefore, \systemname is expected to learn system regularities from historical events and flag deviations accordingly.

\Cref{tab:datasets-table} summarizes the number of training and testing events in each dataset, along with the anomaly ratio in each test set.

\begin{table}
\vspace{-10pt}
\caption{No. train/test events and percentage of anomalies.}
\vspace{-10pt}
\label{tab:datasets-table}
\small
\begin{tabular}{lrrr}
\toprule
\textbf{Dataset} & \#\textbf{Train} & \#\textbf{Test} & \textbf{\% Anomalies} \\
\midrule
Falcon & 2{,}613{,}778 &  277{,}118 & 39.46\% \\
Flask & 2{,}602{,}453 & 483{,}155 & 43.12\%  \\
Live & 2{,}099{,}370 & 485{,}404 & 42.26\% \\
\zrhEvent{} & 450{,}026 & 11{,}269 & 10.84\% \\
BETH & 835{,}250 & 156{,}318 & 5.71\% \\            
\bottomrule
\end{tabular}
\vspace{-1.5em}
\end{table}

\begin{table*}
\vspace{-10pt}
\caption{The anomaly detection performance of \systemname and baselines on five datasets (Falcon, Flask, Live, \zrhEvent{}, and BETH) in terms of Precision, Recall, and F1-score. We report the mean and standard deviation over ten different days/runs of test data. We highlight the best values in \textit{bold} and \underline{underline} the second best.
\vspace{-10pt}
}
\label{tab:results-table}
\resizebox{\textwidth}{!}{%
\setlength\tabcolsep{2pt}
\begin{tabular}{@{}cccccccccccccccc@{}}
\toprule
& \multicolumn{3}{c}{\textbf{Falcon}} & \multicolumn{3}{c}{\textbf{Flask}} & \multicolumn{3}{c}{\textbf{Live}} & \multicolumn{3}{c}{\textbf{\zrhEvent{}}} & \multicolumn{3}{c}{\textbf{BETH}} \\ 
\cmidrule(lr){2-4} \cmidrule(lr){5-7} \cmidrule(lr){8-10} \cmidrule(lr){11-13} \cmidrule(lr){14-16} 
& Precision & Recall & F1-Score & Precision & Recall & F1-Score & Precision & Recall & F1-Score & Precision & Recall & F1-Score & Precision & Recall & F1-Score \\ \midrule
\textbf{\systemname} &\textbf{0.99±0.02} & \textbf{0.96±0.05} & \textbf{0.97±0.03} & \textbf{0.97±0.30} & \textbf{0.94±0.12} & \textbf{0.95±0.26} & \textbf{1.00±0.00} & \textbf{0.99±0.02} & \textbf{0.99±0.01} & \textbf{0.92±0.02} & 0.77±0.00 & 0.84±0.01 & 0.13±0.01 & \textbf{1.00±0.00} & 0.22±0.02 \\
\midrule
DeepSVDD-D & 0.55±0.44 & 0.26±0.34 & 0.31±0.37 & 0.58±0.50 & 0.31±0.36 & 0.37±0.39 & \underline{0.80±0.41} & 0.42±0.39 & \underline{0.48±0.41} & 0.80±0.22 & 0.65±0.22 & 0.66±0.03 & 0.23±0.00 & \textbf{1.00±0.00} & 0.37±0.00 \\
DeepSVDD-T & 0.62±0.45 & 0.39±0.38 & 0.44±0.40 & 0.47±0.46 & 0.34±0.34 & 0.34±0.36 & 0.70±0.44 & 0.37±0.34 & 0.43±0.36 & 0.82±0.10 & 0.81±0.24 & 0.79±0.12 & 0.61±0.45 & \underline{0.95±0.05} & 0.66±0.34 \\
RDP-D & 0.14±0.16 & \underline{0.82±0.39} & 0.22±0.23 & 0.12±0.27 & \underline{0.80±0.41} & 0.15±0.28 & 0.12±0.27 & \underline{0.50±0.51} & 0.14±0.29 & 0.11±0.00 & \textbf{1.00±0.00} & 0.20±0.00 & 0.06±0.00 & \textbf{1.00±0.00} & 0.11±0.00 \\
RDP-T & 0.14±0.16 & 0.80±0.40 & 0.21±0.23 & 0.12±0.26 & \underline{0.80±0.40} & 0.15±0.28 & 0.12±0.27 & \underline{0.50±0.50} & 0.14±0.28 & 0.11±0.00 & \textbf{1.00±0.00} & 0.20±0.00 & 0.06±0.00 & \textbf{1.00±0.00} & 0.11±0.00 \\
RCA-D & 0.40±0.36 & 0.35±0.38 & 0.23±0.24 & 0.34±0.36 & 0.55±0.44 & 0.37±0.34 & 0.35±0.46 & 0.13±0.25 & 0.16±0.30 & 0.43±0.00 & 0.92±0.00 & 0.58±0.00 & 0.70±0.28 & 0.87±0.05 & 0.76±0.19 \\
RCA-T & 0.40±0.39 & 0.37±0.41 & 0.26±0.30 & 0.30±0.37 & 0.46±0.41 & 0.28±0.31 & 0.27±0.37 & 0.32±0.40 & 0.22±0.30 & 0.55±0.21 & 0.62±0.22 & 0.53±0.08 & \underline{0.98±0.00} & 0.92±0.00 & \textbf{0.95±0.00} \\
NeuTraL-D & \underline{0.85±0.33} & 0.46±0.40 & 0.52±0.40 & 0.64±0.48 & 0.55±0.40 & \underline{0.51±0.40} & 0.59±0.48 & 0.29±0.36 & 0.35±0.38 & 0.88±0.00 & \underline{0.99±0.00} & \textbf{0.93±0.00} & \textbf{1.00±0.00} & 0.90±0.00 & \textbf{0.95±0.00} \\
NeuTraL-T & 0.82±0.34 & 0.55±0.39 & \underline{0.60±0.38} & \underline{0.68±0.45} & 0.45±0.38 & 0.49±0.39 & 0.68±0.46 & 0.35±0.36 & 0.41±0.37 & \underline{0.89±0.03} & 0.91±0.21 & 0.89±0.14 & \textbf{1.00±0.00} & 0.91±0.01 & \textbf{0.95±0.00} \\
ICL-D & 0.60±0.48 & 0.22±0.35 & 0.25±0.37 & 0.44±0.46 & 0.52±0.41 & 0.35±0.35 & 0.49±0.48 & 0.35±0.32 & 0.37±0.38 & 0.86±0.00 & \textbf{1.00±0.00} & \underline{0.92±0.00} & 0.55±0.00 & 0.90±0.00 & 0.68±0.00 \\
ICL-T & 0.66±0.44 & 0.39±0.38 & 0.43±0.40 & 0.54±0.49 & 0.36±0.38 & 0.32±0.35 & 0.61±0.49 & 0.22±0.27 & 0.29±0.32 & 0.87±0.05 & \underline{0.99±0.01} & \underline{0.92±0.03} & 0.45±0.44 & \underline{0.95±0.03} & 0.50±0.43 \\
DIF-D & 0.21±0.20 & 0.41±0.48 & 0.15±0.17 & 0.32±0.31 & 0.37±0.28 & 0.30±0.24 & 0.27±0.32 & 0.43±0.44 & 0.26±0.28 & 0.67±0.00 & 0.45±0.00 & 0.54±0.00 & \underline{0.98±0.01} & 0.91±0.01 & \underline{0.94±0.00} \\
DIF-T & 0.15±0.17 & 0.27±0.41 & 0.10±0.15 & 0.21±0.26 & 0.37±0.38 & 0.19±0.20 & 0.18±0.28 & 0.24±0.37 & 0.15±0.23 & 0.43±0.12 & 0.84±0.24 & 0.53±0.03 & 0.96±0.03 & 0.92±0.01 & \underline{0.94±0.01} \\
SLAD-D & 0.62±0.42 & 0.45±0.41 & 0.48±0.40 & 0.53±0.44 & 0.34±0.35 & 0.38±0.36 & 0.51±0.48 & 0.34±0.42 & 0.35±0.43 & 0.49±0.00 & 0.98±0.00 & 0.65±0.00 & \underline{0.98±0.00} & 0.90±0.00 & \underline{0.94±0.00} \\
SLAD-T & 0.36±0.35 & 0.59±0.45 & 0.37±0.33 & 0.35±0.44 & 0.29±0.41 & 0.25±0.37 & 0.23±0.41 & 0.25±0.42 & 0.21±0.37 & 0.79±0.10 & 0.85±0.09 & 0.81±0.05 & 0.94±0.09 & 0.92±0.01 & 0.93±0.04 \\
\bottomrule
\end{tabular}
}
\vspace{-10pt}
\end{table*}

\subsubsection{Evaluation Metrics}

We benchmark \systemname and the baselines for point-wise anomaly detection, where the goal is to determine whether each individual event is normal or anomalous. We use standard evaluation metrics: Precision, Recall, and F1-Score. 
Correctly identifying an abnormal event is counted as a True Positive (TP). Similarly, misclassifying an abnormal event as normal results in a False Negative (FN), while misclassifying a normal event as abnormal results in a False Positive (FP). The formulas for computing Precision, Recall, and F1 Score are as follows:
\begin{equation}
\textit{Pre} = \frac{\textit{TP}}{\textit{TP} + \textit{FP}}, \quad \textit{Rec} = \frac{\textit{TP}}{\textit{TP} + \textit{FN}}, \quad \textit{F1} = \frac{2 \times \textit{Pre} \times \textit{Rec}}{\textit{Pre} + \textit{Rec}}.
\end{equation}

\subsubsection{Baselines} \label{sec:baselines}

We select \numbaseline anomaly detection algorithms from prior studies for performance comparison, namely: DeepSVDD~\cite{deepsvdd}, RDP~\cite{rdp}, RCA~\cite{rca}, NeuTraL~\cite{neutral}, ICL~\cite{icl}, DIF~\cite{dif}, and SLAD~\cite{slad}. For each baseline, we evaluate in two settings: (1) using the default suggested parameters and (2) performing hyperparameter tuning to identify the threshold that produces the best performance. Specifically, for hyperparameter tuning, we use one day of test data for Falcon, Flask, and Live, and the full dataset for \zrhEvent{} and BETH to iterate over a wide range of parameters, including hidden dimensions, training epochs, batch size, learning rate, and anomaly detection threshold, to select the optimal configuration. To ensure reproducibility, we use the implementations provided in DeepOD~\cite{deepod2021}. It is worth noting that all baselines require GPU machines and are prohibitively slow when run on CPU. We conduct all experiments using machines equipped with one T4 GPU, 8 vCPUs, and 32 GiB of memory, but \systemname{} uses a single vCPU.

\subsection{Effectiveness in Anomaly Detection}

\Cref{tab:results-table} presents a comparison of \systemname against \numbaseline anomaly detection baselines across five datasets. For each baseline, the default parameter setting is denoted by the \texttt{-D} postfix, and the tuned parameter setting is denoted by the \texttt{-T} postfix. We report the mean and standard deviation of the evaluation metrics (Precision, Recall, and F1-Score) over 10 different runs and test samples. We obtain the following insights:

\textbf{(1) \systemname is highly effective across different runs and test samples.} \systemname achieves an F1-score greater than 0.95 on the three event-based datasets and maintains very low variance across different runs and test samples. In contrast, all baselines struggle to maintain consistent performance because (1) they depend on random initialization, and (2) different test samples expose different anomaly symptoms, and the baselines detect some anomalies better than others. This is most noticeable on Falcon, Flask, and Live, since each run corresponds to a different test day for these datasets, whereas \zrhEvent{} and BETH are tested on a single~day. 

\textbf{(2) The baselines perform well on \zrhEvent{} and BETH datasets but fail to generalize to complex datasets (Falcon, Flask, Live).} As described in Section~\ref{sec:dataset}, Falcon, Flask, and Live are collected from large cloud systems with thousands of resources and interacting actors. This means that, though there are more events in total, there are fewer events showing the behavior of each individual entity. This reduces the effectiveness of deep learning-based approaches compared to datasets with fewer actors, such as \zrhEvent{} and BETH, for varied reasons. For example, DeepSVDD relies on a deep neural network to extract support vectors from historical event data, which is unreliable for a small number of training events per entity.
\systemname learns patterns that identify entities and model interactions between them, even if only a few events are observed. This contrast is most visible in the Live dataset, where \systemname outperforms the baselines most significantly. This dataset contains interactions between a larger number of actors (meaning fewer events to describe the actions of each actor) and many anomalies resulting from actors performing unseen operations. 

\textbf{(3) Hyperparameter tuning can improve anomaly detection performance in specific cases but fails to generalize across all datasets.} For example, tuning improves DeepSVDD's~\cite{deepsvdd} F1-score from 0.37 to 0.66 on the BETH dataset and from 0.66 to 0.74 on \zrhEvent{}, but fails to improve the performance of RDP~\cite{rdp} and SLAD~\cite{slad} on Falcon. This is understandable, as the parameters are tuned using data from a single day and applied across ten days, and systems exhibit different anomaly symptoms on different days. Similar observations on parameter tuning have been reported in prior research~\cite{pham2024root}. We do not perform any additional tuning on \systemname using the test datasets, it is always run with the same settings, which gives the baselines a more favorable comparison.

\textbf{(4) \systemname underperforms on BETH}. BETH contains numerical attributes associated with low-level interactions (e.g., IP addresses, port numbers) that \systemname's regexes cannot capture well. Since those values are not generalized, any deviation in the test set is marked as anomalous (e.g., a new IP address).
An interesting direction for future work would be to incorporate generalization techniques adapted to other datatypes. For example, IP addresses can be handled by using location and system information, and generalizing over those strings as opposed to the IP address itself.

\subsection{Efficiency of \systemname}

The efficiency of \systemname is critical for its deployment in large-scale systems.  Table~\ref{tab:througput} reports the throughput of \systemname against seven baselines across five datasets. Note that all baselines require a GPU to perform training and inference in useful time; thus, the reported numbers for the baselines are obtained from T4 GPU-equipped~\footnote{https://www.nvidia.com/en-us/data-center/tesla-t4/} machines while \systemname uses only CPU. On a CPU-only environment, \systemname achieves a \textbf{400$\times$} speedup in training time and a \textbf{120$\times$} speedup in inference time compared to NeuTraL, the second-best performing baseline, on the Falcon dataset.
We obtain the following insights:

\textbf{(1) \systemname processes over 10k events per second (\textasciitilde\throughput).} 
\systemname achieves the highest throughput on two datasets (Falcon and Flask) and ranks second on two others (Live and BETH), despite being executed on CPU-only infrastructure. 

\textbf{(2) DeepSVDD achieves high throughput but lacks consistency.} DeepSVDD encodes the values of event entities as one-hot vectors, which enables high throughput on datasets with lower cardinality entities (e.g., Boolean values), such as BETH and \zrhEvent{}. However, its performance degrades significantly on datasets with high-cardinality entities (such as randomized user IDs), as seen in Falcon and Flask, because these result in very high-dimensional one-hot vectors. In contrast, \systemname's inference procedure relies entirely on regex matching, allowing it to maintain high, stable throughput across all datasets.

\textbf{(3) \systemname achieves a better effectiveness–efficiency trade-off than NeuTraL.} As shown in Table~\ref{tab:results-table}, NeuTraL delivers anomaly detection effectiveness comparable to \systemname. However, it lags significantly in efficiency. For instance, on the Falcon and Flask datasets, \systemname (CPU) is 8\(\times\) faster than NeuTraL (GPU).

\begin{table}[t]
\caption{Throughput (in thousands of events per second) of \systemname{}(CPU) and baseline methods (GPU) across five datasets. Each value represents the mean ± standard deviation. We \textit{bold} the best values and \underline{underline} the second best.} \label{tab:througput}
\vspace{-5pt}
\setlength{\tabcolsep}{2pt}
\small\begin{tabular}{lccccc}
\toprule
\textbf{Method} & \textbf{Falcon} & \textbf{Flask} & \textbf{Live} & \textbf{\zrhEvent{}} & \textbf{BETH} \\
\midrule
DeepSVDD & \underline{9.89±0.2} & \underline{8.13±0.4} & \underline{12.90±0.2} & \textbf{61.56±1.7} & \textbf{98.35±1.2} \\
RDP & 6.17±2.8 & 3.99±0.1 & 3.67±0.0 & \underline{34.33±0.2} & \underline{50.36±0.5} \\
RCA & 0.58±0.0 & 0.66±0.5 & 0.41±0.0 & 4.29±0.1 & 5.81±0.1 \\
NeuTraL & 2.11±0.0 & 1.87±0.0 & 3.76±1.9 & 16.63±0.5 & 21.04±0.4 \\
ICL & 2.50±0.1 & 1.26±1.0 & 0.69±0.3 & 16.49±0.2 & 26.66±1.2 \\
DIF & 0.40±0.0 & 0.68±0.0 & 0.63±0.0 & 0.75±0.0 & 0.76±0.0 \\
SLAD & 0.13±0.0 & 0.08±0.0 & 0.07±0.0 & 0.96±0.1 & 1.57±0.0 \\
\midrule
\textbf{\systemname} & \textbf{18.18±0.8} & \textbf{18.37±0.6} & \textbf{17.80±0.6} & 16.60±0.1 & 31.63 ± 0.8  \\
\bottomrule
\end{tabular} 
\vspace{-10pt}
\end{table}

\subsection{Sensitivity and Robustness of \systemname} \label{sec:ablation}

In this section, we first evaluate the sensitivity of \systemname to different parameter settings (similarity threshold, etc). Second, we evaluate the robustness of \systemname to noise in the training set (drop, duplicate, shuffle), demonstrating that \systemname is a robust and practical anomaly detection method.

\subsubsection{Sensitivity to different parameter settings}

\begin{figure}
    \centering
    \includegraphics[width=.92\linewidth]{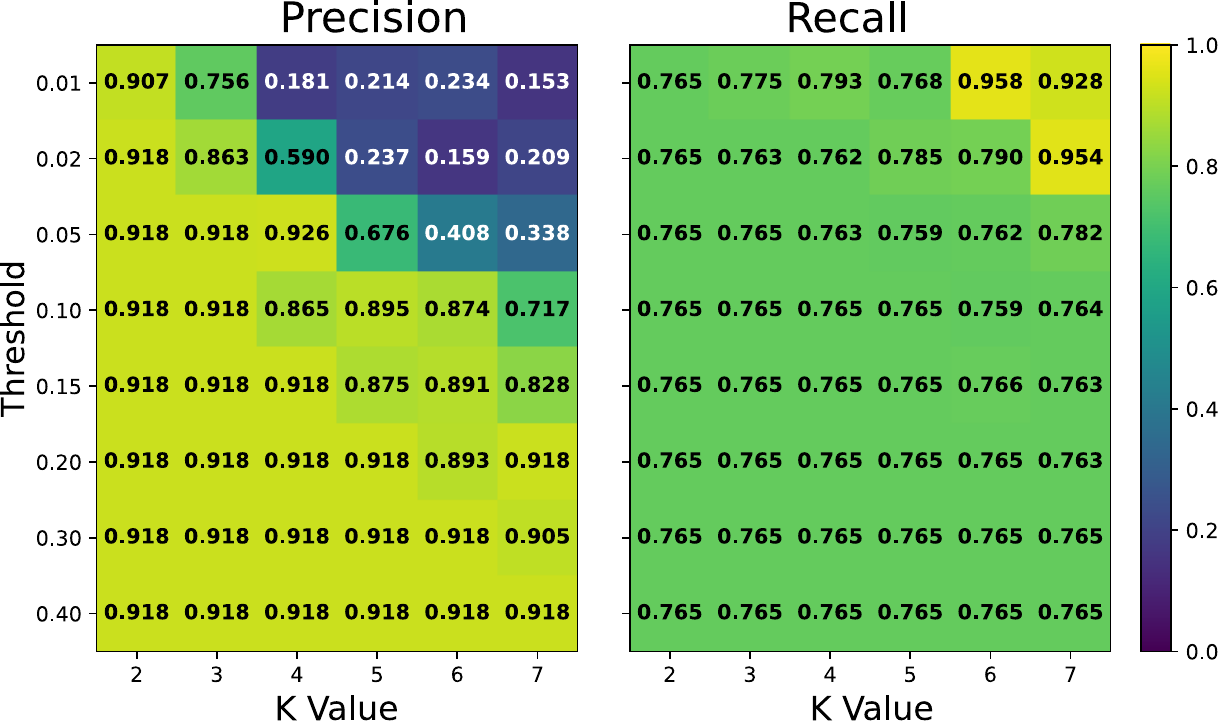}
    \vspace{-10pt}
    \caption{Sensitivity analysis of the similarity score merge threshold and value \(k\) in \Cref{eq:lsimrank-iter} on the Out dataset.}
    \vspace{-10pt}
    \label{fig:sensitivity}
\end{figure}

We test \systemname with different values for \(k\) in \Cref{eq:lsimrank-iter}, and the threshold above which we merge two vertices (as explained in \Cref{sec:graph-sim}). Higher \(k\) values mean more iterations when computing the similarity between pairs of vertices in the hypergraph. 
A higher threshold for merging vertices means that fewer vertices will be merged in each iteration of \systemname, leaving fewer opportunities for generalization beyond the observed events.
\Cref{fig:sensitivity} shows the resulting precision and recall for the \zrhEvent{} data set. The results for the remaining datasets are in \Cref{extra:sensitivity}. In general, we observe that \textbf{higher values of \(k\) can help recall (generalization is more specific) but hurt precision}. The higher \(k\) is, the higher the threshold needs to be to obtain good precision.
The unexpected result comes from the low precision for low threshold values. We would expect the patterns to generalize more, since more attempts to merge will be made, but this results in many more failures to merge dissimilar entities. This highlights a limit (fortunate in this case) of the regex synthesis approach, which cannot find a solution when many different strings need to be captured by a single pattern.

\subsubsection{Robustness to noise in training set}

\begin{figure}
\centering
\includegraphics[width=0.9\linewidth]{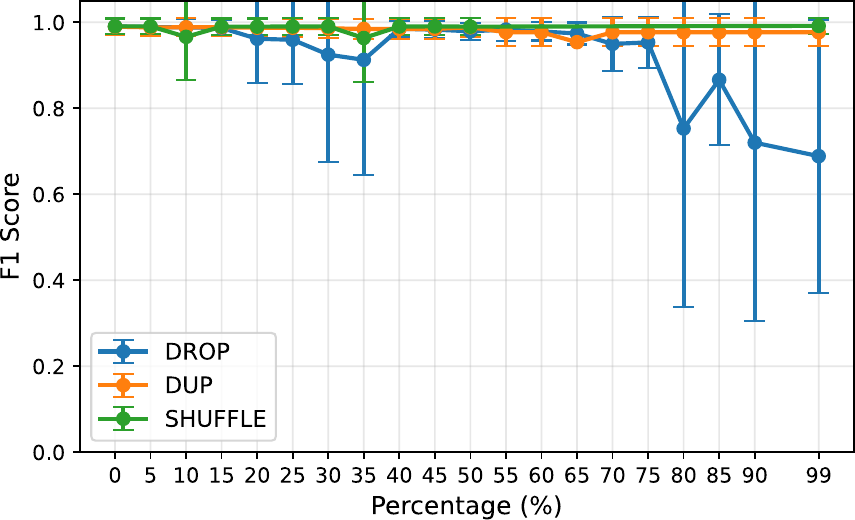}
\vspace{-10pt}
\caption{The robustness of \systemname w.r.t different noise types and levels on the Live dataset.}
\label{fig:rob}
\vspace{-10pt}
\end{figure}

We show that our method is robust to noisy training data by randomly dropping, duplicating, or shuffling up to 99\% of the total events in the training data. This allows us to evaluate how \systemname responds to different types of noise.
\textbf{\systemname{} achieves perfect recall regardless of the noise conditions introduced}. \systemname consistently detects all true anomalies regardless of how noisy the training data is. The F1 score is relatively constant across all levels of noise for shuffling and duplicating: this is expected, as \systemname does not take into account the ordering between events or their frequency. Our approach ingests every event in a streaming manner and remembers every single event.
The precision degrades more noticeably when training events are \textit{dropped}, but we need to drop 70\% of events for the F1 score to drop below 90\% on the Live dataset. However, dropping a single event might cause imprecision because a pattern could not be learned. This explains the high variances of F1 scores in \Cref{fig:rob}.

%% file: sections/conclusion.tex
\section{Conclusion and Future Work}\label{sec:conclusion}

\systemname is a novel approach that learns an interpretable model of normal system behavior from events by leveraging observable relations between entities, and learning patterns over similar entities. This model can be used as an efficient and interpretable anomaly detection filter. The two main components of our system are a method to measure similarity between labeled vertices in a hypergraph, and a technique to generalize over the labels in the hypergraph. We believe that we can improve over those two components in order to make this approach more applicable to other types of logs. Our only instantiation of the label generalization is our proposed regex synthesis technique, but we could use other approaches to generalize over semantically rich datatypes such as IP addresses.

\balance

%% file: appendix/extra.tex



\section{Extra}\label{sec:extra}

\paragraph{Fully Merging Vertices}
Two vertices $v_1 = (k_1, r_1, \tau_1)$ and $v_2 = (k_2, r_2, \tau_2)$ in a hypergraph $\Graph$ can be merged together if:
\begin{itemize}
  \item $k_1 = k_2$ and $\tau_1 = \tau_2$, that is, they have the same type and key.
  \item $\text{MergeRegex}(\{r_1, r_2\}, R^-) \neq \emptyset$, that is, their values can be captured by the same regex.
  \item $\text{SimScore}(v_1, v_2, \Graph) > c$ where $c$ is some predefined threshold for similarity.
\end{itemize}

If two vertices can be merged, then the graph is updated by removing vertices $v_1$ and $v_2$, and replacing $v_1$ and $v_2$ in the edges by a new vertex $v_3 = (k_1, \text{MergeRegex}(\{r_1, r_2\},R^-), \tau_1)$. Duplicate edges are merged together.

\paragraph{Partially merging vertices}
Sometimes $\text{SimScore}(v_1, v_2, \Graph) \leq c$ but there exists a subgraph $H \subset \Graph$ such that $\text{SimScore}(v_1, v_2, H) > c$.
A new vertex $v_3 = (k_1, \text{MergeRegex}(\{r_1, r_2\}, R^-), \tau_1)$ is added to the graph, and $v_1$ and $v_2$ are replaced by $v_3$ in the edges in $H$. The edges in $\Graph \setminus H$ are unchanged. Intuitively, the vertices are similar only if some of the hyperedges are considered.


\begin{proposition}[Edge Uniqueness and Event Matching]
  The edges in $\Edges$ are unique iff for any given event $\eta = \{(k_i, r_i, \tau_i)\}_{1 \leq i \leq n}$, there is at most one edge $e \in \Edges$ such that 
 \begin{itemize}
   \item $e$ has the same signature as $\eta$ , i.e. there exists values $s_1, \ldots , s_n$ such that $e = \{(k_i, s_i, \tau_i)\}_{1 \leq i \leq n}$
   \item the values match, i.e. $\forall 1 \leq i \leq n \cdot s_i \simeq r_i$.
 \end{itemize}
\end{proposition}

%% file: appendix/add-eval.tex
\section{Additional Evaluation Results}

\subsection{Parameter Sensitivity}\label{extra:sensitivity}

\begin{figure*}
\centering
\includegraphics[width=.95\linewidth]{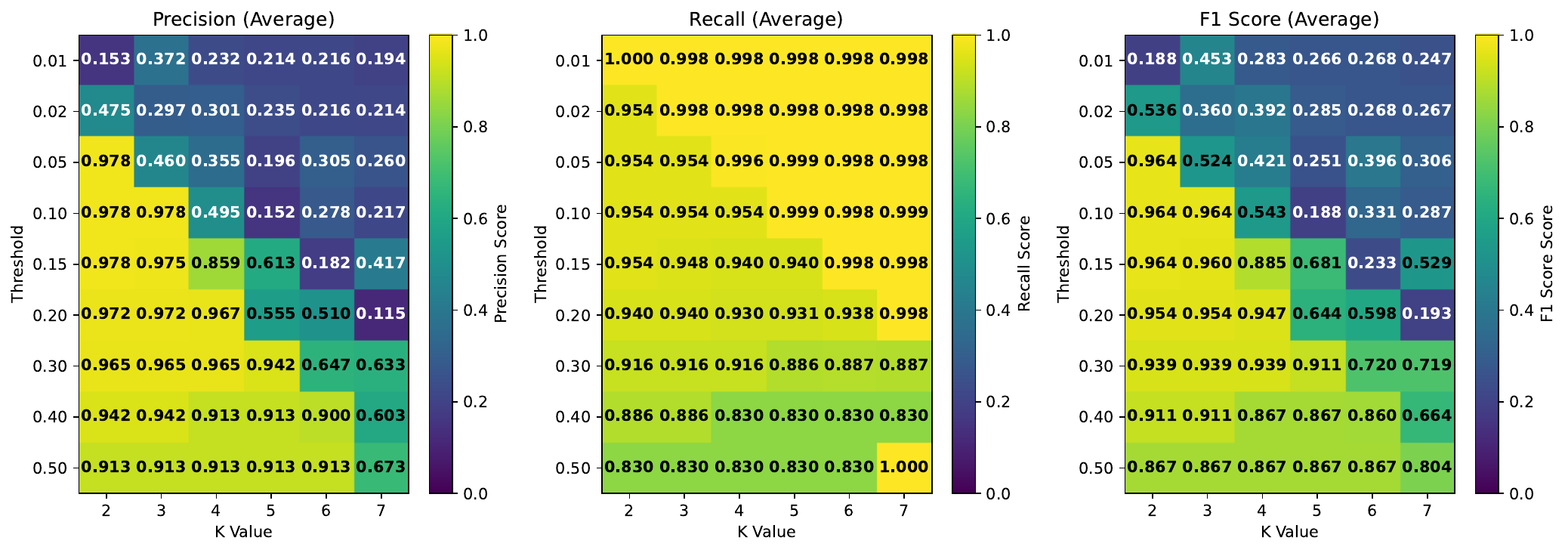}
\caption{Sensitivity analysis of the similarity score merge threshold and value \(k\) in \Cref{eq:lsimrank-iter} on the Flask dataset.}
\label{fig:sensitivity-flask}
\end{figure*}

\begin{figure*}
\centering
\includegraphics[width=.95\linewidth]{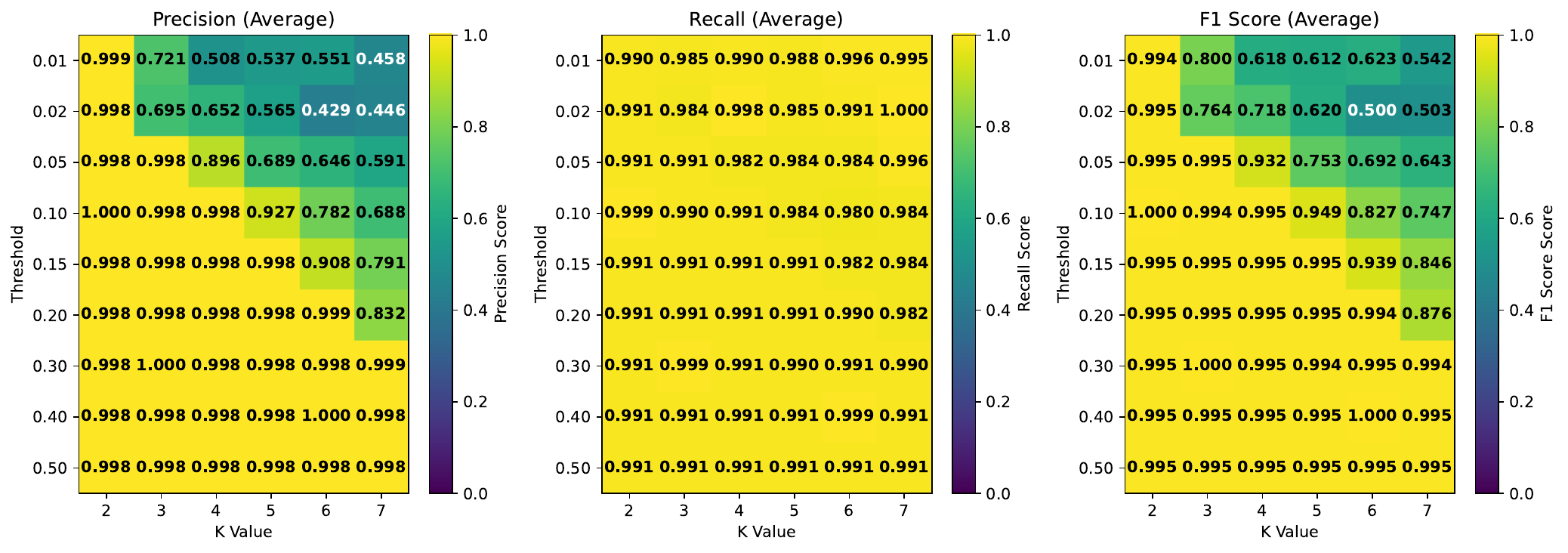}
\caption{Sensitivity analysis of the similarity score merge threshold and value \(k\) in \Cref{eq:lsimrank-iter} on the Live dataset.}
\label{fig:sensitivity-live}
\end{figure*}

\begin{figure*}
\centering
\includegraphics[width=.95\linewidth]{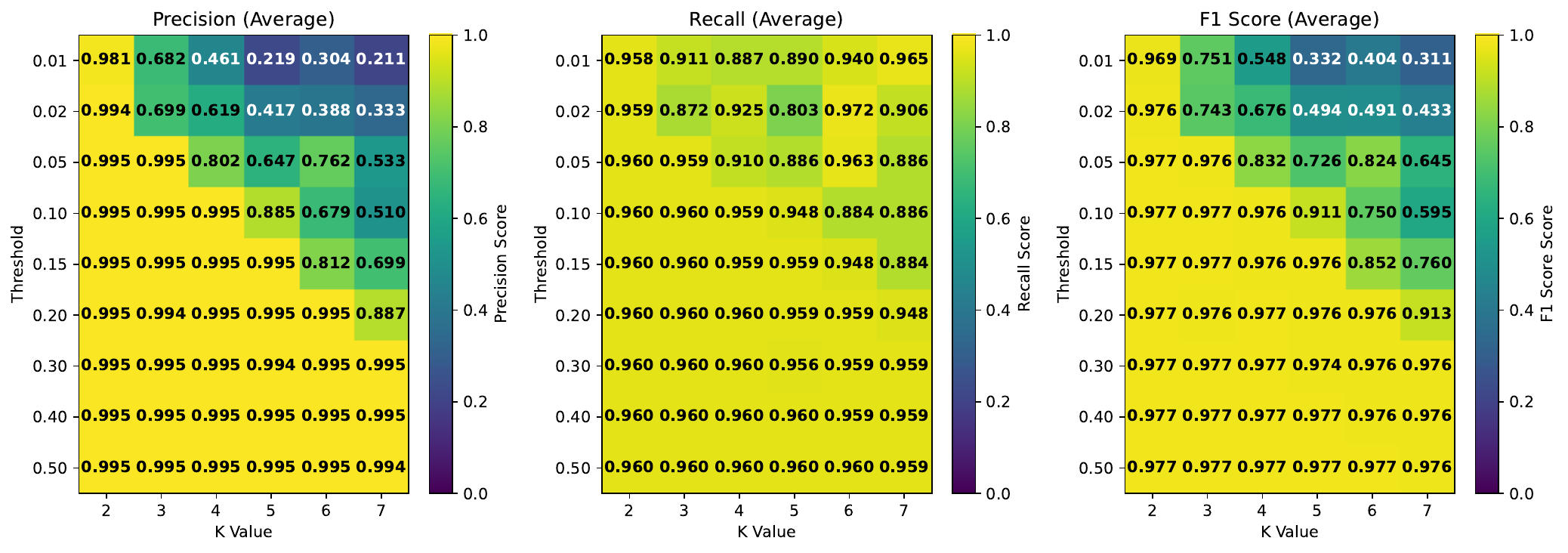}
\caption{Sensitivity analysis of the similarity score merge threshold and value \(k\) in \Cref{eq:lsimrank-iter} on the Falcon dataset.} \label{fig:sensitivity-falcon}
\end{figure*}

\Cref{fig:sensitivity-flask,fig:sensitivity-live,fig:sensitivity-falcon} show the results of the sensitivity analysis for the Falcon, Flask, and Live datasets.

